\documentclass[11pt,twoside]{article}
\usepackage{asp2006}
\usepackage{lscape}
\usepackage{epsfig,graphicx,natbib,url}  
\begin{document}

\markboth{K.P. Reardon et al.}{Service-Mode Observations for Ground-Based Solar Physics}
\title{Service-Mode Observations for Ground-Based \\
         Solar Physics}  
\author{Kevin P. Reardon$^{1,2}$, 
	Thomas Rimmele$^{2}$, 
	Alexandra Tritschler$^{2}$, 
	Gianna Cauzzi$^{1}$, 
	Friedrich Woeger$^{2}$, 
	Han Uitenbroek$^{2}$, 
	Saku Tsuneta$^{3}$
	and
	Thomas Berger$^{4}$}
\affil{$^1$ INAF/Arcetri Astrophysical Observatory, Florence, Italy\\
       $^2$ National Solar Observatory, Sunspot, NM, USA\\
       $^3$ National Astronomical Observatory of Japan, Mitaka, Japan\\
       $^4$ Lockheed Martin Solar and Astrophysics Laboratory, Palo Alto, USA}
  
\begin{abstract} 
There are significant advantages in combining {\it Hinode} observations with ground-based instruments 
that can observe additional spectral diagnostics at higher data rates and with greater flexibility. 
However, ground-based observations, because of the random effects of weather and seeing as well as 
the complexities data analysis due to changing instrumental configurations, have traditionally been 
less efficient than satellite observations in producing useful datasets. Future large ground-based 
telescopes will need to find new ways to optimize both their operational efficiency and scientific output.

We have begun experimenting with service-mode or queue-mode observations at the Dunn Solar 
Telescope using the Interferometric Bidimensional Spectrometer (IBIS) as part of joint {\it Hinode} 
campaigns. We describe our experiences and the advantages of such an observing mode for solar 
physics.
\end{abstract}

\section{Ground-Based Solar Observations}      \label{reardon-sec:traditional}
Even with the many important space-based observatories, ground-based instruments will continue to 
provide unique observations unavailable from space, primarily because of the availability of higher 
spatial resolution or light collecting power, the much greater data rates that can be achieved, and the 
higher level of flexibility, for example in real-time targeting or instrument upgrades.

High-resolution solar observations are characterized by observing programs full of tradeoffs that are made 
based on the specific scientific goals\footnote{
While ground-based observations come in many forms, we concentrate on high-resolution 
observations in the visible and infrared regimes since there are many prominent facilities catering to 
such usage and they constitute a significant portion of the expenditures for ground-based solar physics
}.
Limitations on instrumentation, detectors, terrestrial atmosphere, and photon flux require decisions on 
observed wavelengths, image scale, cadence, and coverage, as examples.  Since high-resolution 
observations have limited fields of view, the particular target on the solar surface must be selected, 
taking into account both the rotation of the Sun as well as the overall evolution of structures on 
timescales of hours. 

Principal investigators submit observing proposals that describe their scientific goals and their choices 
for the tradeoffs outlined above. They also specify the type of solar structures required for their 
observing program. Successful proposals are generally assigned fixed periods of one to two weeks at 
the telescope, usually many months in advance. A portion of the ``observing'' time is dedicated to 
instrumental setup particular the specific program, often because major modifications to the 
instrumental configuration are desired (and allowed). For a variety of reasons, one or more of the 
proposers are typically present at the telescope during the full observing run. Often they are required to 
actually setup the instrumentation and perform observational tasks, or at least to make further choices 
on the observational setup in near real-time. They also perform immediate quality assurance, both in 
terms of monitoring the immediate conditions ({\it i.e.} the ``seeing'') and checking the acquired data. 
Finally, given the rapid variability of the Sun, often the selection of the target requires direct 
visualization of the structures present at the time of the observations.

This mode of observation, commonly referred to as ``classical-mode'' or ``visitor-mode'' in astronomy, 
has long been the standard in solar physics. This laboratory concept, allowing scientists to 
manipulate instruments to construct {\it ad hoc} configurations, indeed offers a high level of flexibility in 
the instrumental setup and observational sequences, while providing for straightforward planning of 
telescope operations. It allows selected scientists to (eventually) become proficient in instrumental and 
observational techniques.

Clearly, however, there are many drawbacks to such an approach. The investigators must invest time 
and money traveling to the observatory and must labor to become proficient in the operation and 
intricacies of the instrumentation. They are also expected to perform their own data reduction with 
limited software packages and documentation. All this is for an uncertain gain given the vagaries of the 
weather and solar conditions. Time is lost changing setups between programs, and there is little 
opportunity to adapt the observational schedule to the immediate conditions. Only a few dozen 
observing programs can be carried out in a year, and even then the chance of successful execution 
remains low. All these limitations tightly constrain the scientific output of ground-based telescopes to 
just 10--20 papers per year.

\section{Service-Mode Observations}      \label{reardon-sec:servicemode}

Experience at ground-based nighttime facilities has shown that new approaches, such as 
service-mode observing, can offer significant gains in observational efficiency and scientific output. 
Service-mode operations, also known as queue or dynamic scheduling, are based on the definition of a 
prioritized list of observing programs by an allocation committee, and then the selection by telescope 
staff of the appropriate program to be run given the actual observing conditions 
\citep{Silva:2002p11470}. Both VLT \citep{Comeron:2006p11421} and 
Gemini \citep{Puxley:2006p11403} perform 
a significant amount of their observations in such a mode (50\% and 90\%, respectively) and future 
facilities like ALMA will be operated exclusively in this mode. Given the dynamic nature of the 
scheduling, investigators are typically not present at the telescope when their observing program is 
executed.

The transition from classical to service-mode observations undertaken in nighttime astronomy during 
the 90's was driven by the need to improve the efficiency and scientific output of new large facilities with 
high total costs \citep{Robson:1996p11354}. In fact, the flexible scheduling not only allows more 
programs to be run, but also ensures that the highest ranked proposals have the best chance of being 
completed. An essential advantage is that it improves the odds that the moments of above-average 
atmospheric conditions are used for programs requiring the best conditions. This mode also facilitates 
observations of targets of opportunity and programs requiring observations spread over many weeks or 
months.

Solar physics is facing a very similar situation as it moves toward the construction of large-aperture 
solar telescopes such as ATST and EST. These facilities will be expensive and will require 
commensurate annual operational budgets. With costs comparable to current 8--10 meter telescopes, 
they will be expected to achieve a similar level of scientific output, namely resulting in upwards of 
50--100 papers published per year using data from the telescope. To achieve this goal it will be crucial 
to increase the number of observing programs that are run each year in concert with a broadening the 
currently limited user-base for ground-based solar observations.

Solar physics would be well served by the flexible scheduling of service-mode observations. The 
success of many scientific programs requires simultaneously satisfying at least two conditions that both 
have significant temporal variability\,---\,acceptable atmospheric conditions and suitable solar features. 
Achieving the necessary intersection of these two parameters is arguably more critical than the 
presence of the investigator at the telescope. If the observing program has been well defined in 
advance, one remaining element possibly requiring investigator input at the time of the observations is 
the selection of the target. This is different from many nighttime observations because the structures on 
the solar surface are rapidly evolving and high-resolution instruments tend to have small 
fields-of-view that cover only a portion of features such as active regions or filaments. The 
presence of a resident astronomer at the telescope and the possibility of real-time feedback from 
the investigator may nonetheless allow for reliable choice of target in the majority of cases.

As an additional advantage, the ability to switch observing programs as the seeing conditions vary 
during the day, for example performing coronal or IR observations in the afternoon, would allow for an 
efficient usage of the sunlight hours. It will be imperative to develop science programs that can be 
productively carried out even during periods of ``sub-optimal'' seeing conditions. Long-term programs, 
such as high-resolution synoptic observations, would also be enabled by dynamic scheduling.

Supporting service-mode operations requires an extensive support infrastructure throughout the facility, 
extending from the observation planning and program execution through data processing and delivery. 
In a planning phase, investigators will need to provide a a full description of their program requirements 
and the desired observational setup. Mechanisms must be put in place to smoothly manage the 
experiment queue. Additional or more highly trained staff may be needed to support such an 
approach.
There are certainly additional costs involved with the implementation of such a system. However, given 
the significant baseline operation expenses, improvements in the efficiency of the facility will probably 
be well worth their incremental costs.

This mode of operation also places certain demands on the instrumentation. It must be possible to 
adapt to different observing programs within a short period of time. The calibration sequences must be 
well defined and able to be shared, if possible, among different observing programs. Having common 
data acquisition procedures allows for the development of a standard reduction pipeline able to 
produce usable data for a wide range of potential users.

Some of the greatest problems associated with the transition to service-mode scheduling may not be 
technical, but cultural \citep{Padman:1996p11358}. Current users of solar telescopes will have to be 
satisfied at times with less hands-on time at the telescope. Members of the community who don't 
currently use ground-based data need to be encouraged to become ``customers'' of these facilities. 
Open access to ground-based data, as is standard with many satellites, can also aid in increasing the 
usage of the facility. The more uniform approach to data acquisition and calibration inherent in service-
mode observations can greatly improve the ability to effectively share acquired data among different 
researchers.

\section{Service-Mode Tests with {\it Hinode}}      \label{reardon-sec:hinode}

Spacecraft, by their nature, typically operate using a dynamically scheduled approach. This is indeed 
the case with the {\it Hinode} satellite \citep{Kosugi:2007p6157}, which is also notable because the 
SOT has a suite of instruments with similar capabilities to those of high-resolution ground-based 
telescopes \citep{Tsuneta:2008p5059}. These instruments have a broad group of users from a variety 
of backgrounds and a high rate of scientific output
(we note that {\it Hinode} data analysis is greatly simplified by not having to contend with the variability 
of the terrestrial atmosphere). 
This serves as an indication of the viability of using a similar scheduling approach for ground-based 
telescopes.

The Interferometric Bidimensional Spectrometer \citep[IBIS;][]{Cavallini:2006p5571} installed at the 
Dunn 
Solar Telescope (DST) appears well suited to support service-mode type observations. The instrument 
can be tailored to different observational programs with simple changes of configuration files and 
software controls, without any hardware modification. The characteristics of IBIS are well calibrated and 
the resident observers are able to efficiently operate the instrument even in the absence of the 
investigators. Finally, the data reduction process is well understood and a standard data reduction 
pipeline is under development. Previous tests using a queue of backup observing programs to be run 
during otherwise unused time have also been carried out previously using the Diffraction-Limited 
Spectropolarimeter \citep[DLSP; ][]{2004SPIE.5171..207S}.

Because the IBIS imaging spectroscopy of the chromospheric H$\alpha$ and Ca II 854.2 nm lines is 
highly complementary to the {\it Hinode} data, the service-mode observations were scheduled
as part of two coordinated campaigns in April  and July 2008.
During these periods, five separate observing programs were run, 
with the different programs being chosen on the basis of the solar features present.

The principal investigators provided requests for the desired spectral lines and 
sampling, but none were present at the telescope. Ground-based observations are 
subject to variable seeing conditions, so not all of these datasets were of acceptable 
quality. Several of the better datasets have been fully reduced and provided to the investigators 
and the results from these observations are expected to be published in the future.

\section{Conclusions}      \label{reardon-sec:conclusions}

We have shown that service-mode observations at a ground-based telescope, given instruments well 
adapted to such use, are feasible and may be a more efficient means to provide some users with high-
resolution data. Many scientists may not have the time or observational experience to commit to an 
observing run at a ground-based telescope within the traditional framework. Service-mode scheduling
gives more highly rated programs a better chance of being executed. Such dynamic scheduling would 
ensure that programs are only run when their pre-defined conditions, both in the terrestrial and solar 
atmosphere, are satisfied. 

We believe such an approach will be crucial for future large-aperture solar telescopes such as ATST 
and EST. The systems needed to support such a scheduling approach are best designed as an 
integrated part of such an observatory. There is still much to be learned on the best ways to schedule 
and perform such observations. Some can be drawn from the nighttime practices, but there are many 
aspects that are particular to solar observations which require further experimentation. It is not clear, for 
example, how involved the investigator needs to be in the detailed choice of the target at the time of 
observation. 

Experience at nighttime facilities has shown that it takes some time for a community to become familiar 
with the advantages and tradeoffs that are inherent in service-mode observing but in the end this 
approach is generally broadly endorsed \citep{Comeron:2006p11421}. By introducing this model to the 
community now, its wider application in the initial operational phases of these new facilities may 
proceed more smoothly. All of this argues for continued trials of service-mode scheduling at existing 
solar facilities.




\end{document}